\documentclass[graybox]{svmult}
\usepackage{type1cm}        % activate if the above 3 fonts are
\usepackage{makeidx}         % allows index generation
\usepackage{graphicx}        % standard LaTeX graphics tool
\usepackage{multicol}        % used for the two-column index
\usepackage[bottom]{footmisc}% places footnotes at page bottom

\usepackage{newtxtext}       % 
\usepackage[varvw]{newtxmath}       % selects Times Roman as basic font

\makeindex             % used for the subject index
\usepackage{orcidlink}
\usepackage{subcaption}
\usepackage{fontawesome}
\usepackage{bookmark}
\usepackage{hyperref}
\begin{document}

\title*{Robust Training with Data Augmentation for Medical Imaging Classification\thanks{This paper was accepted to the $9^{\text{th}}$ International Workshop on Health Intelligence (W3PHIAI-25) at AAAI 2025.}}

\author{Josué Martínez-Martínez~\orcidlink{0000-0002-2903-9297}, Olivia Brown~\orcidlink{0009-0006-1010-100X}, Mostafa Karami~\orcidlink{0009-0009-1083-2138}, and Sheida Nabavi~\orcidlink{0000-0002-5996-1020}}
\authorrunning{Josué Martínez-Martínez et al.}
\institute{Josué Martínez-Martínez\textsuperscript{\faEnvelopeO} \at University of Connecticut, 371 Fairfield Way, Storrs, Connecticut, USA, \email{josue.martinez-martinez@uconn.edu}
\and Olivia Brown \at MIT Lincoln Laboratory, 244 Wood Street, Lexington, Massachusetts, USA, \email{olivia.brown@ll.mit.edu} \and Mostafa Karami \at 
University of Connecticut, 371 Fairfield Way, Storrs, Connecticut, USA, \email{mostafa.karami@uconn.edu} \and Sheida Nabavi \at University of Connecticut, 371 Fairfield Way, Storrs, Connecticut, USA, \email{sheida.nabavi@uconn.edu}}
\maketitle
%\footnote{This paper was accepted to the $9^{th}$ International Workshop on Health Intelligence (W3PHIAI-25) at AAAI 2025.}
\abstract*{Each chapter should be preceded by an abstract (no more than 200 words) that summarizes the content. The abstract will appear \textit{online} at \url{www.SpringerLink.com} and be available with unrestricted access. This allows unregistered users to read the abstract as a teaser for the complete chapter.
Please use the 'starred' version of the \texttt{abstract} command for typesetting the text of the online abstracts (cf. source file of this chapter template \texttt{abstract}) and include them with the source files of your manuscript. Use the plain \texttt{abstract} command if the abstract is also to appear in the printed version of the book.}

\abstract{Deep neural networks are increasingly being used to detect and diagnose medical conditions using medical imaging. Despite their utility, these models are highly vulnerable to adversarial attacks and distribution shifts, which can affect diagnostic reliability and undermine trust among healthcare professionals. 
In this study, we propose a robust training algorithm with data augmentation (RTDA) to mitigate these vulnerabilities in medical image classification.
We benchmark classifier robustness against adversarial perturbations and natural variations of RTDA and six competing baseline techniques, including adversarial training and data augmentation approaches in isolation and combination,
using experimental data sets with three different imaging technologies (mammograms, X-rays, and ultrasound).
We demonstrate that RTDA achieves superior robustness against adversarial attacks and improved generalization performance in the presence of distribution shift in each image classification task while maintaining high clean accuracy.
}
\keywords{Adversarial training · Medical images · Distribution shift · Robustness · Data Augmentation}

\section{Introduction}
\label{sec:1}
Deep learning (DL) has shown significant promise in medical imaging applications, especially in the detection and diagnosis of conditions such as breast cancer through mammogram analysis~\cite{abdelhafiz2019deep}.
Despite their increasing adoption, deep neural networks (DNNs) remain vulnerable to adversarial attacks and natural variations.
A meta-analysis of 516 published research works showed that less than 6\% of AI applications for the diagnostic analysis of medical images incorporated external validation, such as distribution shift tests~\cite{kim2019design}.
Furthermore, according to the FUTURE AI guidelines~\cite{karimfutureai}, improving the generalizability and robustness of computer-aided diagnostic systems is essential, particularly in healthcare applications where model safety and reliability are paramount. 

These vulnerabilities are particularly concerning in critical applications such as healthcare, where errors can have profound consequences for patient outcomes.
Adversarial attacks refer to carefully crafted perturbations designed to manipulate the predictions of the model. 
However, distribution shift is the consequence of natural variations caused by differences in imaging equipment, demographics, and scanning protocols between institutions or regions. 
Simultaneously mitigating the effects of adversarial attacks and distribution shift is crucial to guaranteeing the reliability and safety of models in various clinical environments.

Despite the importance of addressing these challenges, robust model development for medical imaging has remained largely focused on either adversarial robustness or distribution shift generalization in isolation. 
This has resulted in models that may withstand one type of perturbation but not another, thereby limiting their utility in real-world clinical environments. In this research, we address this gap by presenting a modified robust training approach that integrates adversarial training and data augmentations. 
Our method, robust training with data augmentation (RTDA), is specifically designed to balance these robustness challenges by incorporating data augmentations that simulate natural variations alongside adversarial training techniques, enabling improved reliability of DL models in medical imaging. 

\section{Background}
\label{sec:2}
% Always give a unique label
% and use \ref{<label>} for cross-references
% and \cite{<label>} for bibliographic references
% use \sectionmark{}
% to alter or adjust the section heading in the running head
\subsection{Adversarial Training}
Adversarial training (AT) was initially introduced as a defense mechanism to counter adversarial attacks, which are carefully constructed perturbations that can lead a model to incorrect predictions~\cite{goodfellow2014explaining}. 
In medical imaging, studies have shown that adversarial attacks can be especially problematic, often requiring only minimal perturbations to deceive a model~\cite{ma2021understanding}. 
Given the potential consequences of such vulnerabilities, adversarial robustness has become a key area of focus in healthcare applications. 
Various AT techniques have been explored, such as TRadeoff-inspired Adversarial DEfense (TRADES) \cite{zhang2019theoretically}, and some methods generate adversarial examples during training to increase the resilience of the model~\cite{madry2017towards}.

However, these methods usually sacrifice clean performance to achieve adversarial robustness.
Recent research has studied adversarial defenses in the medical domain, with techniques such as unsupervised detection of adversarial attacks in high-resolution medical images~\cite{li2020robust} and denoiser-based approaches that protect DNNs from perturbations~\cite{kansal2022defending}. Although effective, these methods do not guarantee adversarial robustness in cases where the method is unable to detect or remove perturbations.

\subsection{Data Augmentation}
Data augmentation has emerged as a practical approach to improve the generalization of DL models, especially in data-scarce domains like medical imaging.
Augmentations such as random flips, rotations, and crops have traditionally been used to increase data diversity, thereby improving model performance~\cite{deepResidual2016}. More advanced augmentation techniques, including PixMix~\cite{hendrycks2022robustness}, and MixUp~\cite{mixup2017}, have been developed to further improve the robustness to natural variations by introducing novel variations in training data. 
Subsequently, AugMix improved the MixUp algorithm by enforcing the consistency between the model predictions of images generated with a variety of augmentations~\cite{hendrycks2019augmix}.
AugMix achieves this by combining a chain of simple augmentation operations and incorporating a consistency loss based on Jensen-Shannon divergence, allowing the model to learn invariant features. 
In the medical domain, an AugMix-inspired method is presented that mixes two data augmentation approaches to generalize to other domains~\cite{garrucho2022domain}.
Later, \cite{li2023domain} proposed a contrastive learning approach to enhance the generalization of deep learning models in computer-aided mammography diagnosis using a multi-style and multi-view unsupervised self-learning scheme to address the challenge of training with diverse data from various scanner vendors.
However, while AugMix and similar methods effectively achieve generalization to other distributions and robustness to common perturbations, they do not explicitly account for adversarial robustness. 

\subsection{Robust Training with Data Augmentation}
The authors~\cite{martinezrobustaugmix}, interested in the idea of achieving robustness to natural and adversarial perturbations, introduced RobustAugmix, a technique that integrates AugMix data augmentation with adversarial training leveraging a consistency loss between clean, augmented, and adversarial examples. 
In their study, the proposed approach achieved robustness to a wide variety of natural perturbations presented in CIFAR-10C~\cite{hendrycks2019benchmarkingneuralnetworkrobustness} while maintaining decent adversarial robustness. 
Later, the authors in~\cite{martinez2023robust} showed that RobustAugMix maintains robustness to common noise without sacrificing clean performance, but at the expense of adversarial accuracy.
In addition, the method was not tested against other types of natural variation that can occur in medical imaging scenarios, such as an increase or decrease in contrast that can cause a distribution shift.

\section{Methodology}
Our objective in this study is to examine the performance of various robust learning techniques under adversarial attacks and distribution shift to find their vulnerabilities, and propose a modification to achieve a more robust model that is capable of withstanding these data variations. In this section, we will first describe the different learning techniques building up to our proposed approach that is designed based on the strengths of the following methodologies:

\subsection{Empirical Risk Minimization (ERM)}
ERM~\cite{vapnik1999overview} is a fundamental principle in machine learning, focusing on minimizing the average error within a training dataset. In ERM, the model parameters are iteratively adjusted to reduce the discrepancy between the predicted outcomes and the actual labels:
\begin{equation} \label{eq:erm}
\min_{\theta} E_{(x,y)\sim D} [L(f_\theta(x),y)].
\end{equation}
The main objective, represented by Equation~\ref{eq:erm}, involves minimizing the expected loss in the data set, where $E_{(x,y)\sim D}$ denotes the expectation in the distribution $D$ of input-output pairs $(x,y)$. Here, $L(\cdot)$ signifies a chosen loss function that quantifies the disparity between the predictions made by the model $f_\theta(x)$ and the actual labels $y$. The optimization process adjusts the parameters $\theta$ of the model, often a neural network, to enhance its predictive accuracy in the training data. ERM serves as a cornerstone in model training, laying the groundwork for various machine learning algorithms across diverse domains.

\subsection{Adversarial Training (AT)}
\cite{madry2017towards} formalized the adversarial training objective as a form of robust optimization:
\begin{equation} \label{eq:robust}
\min_{\theta} E_{(x,y)\sim D} [\max_{||\delta||_p<\epsilon}L(f_\theta(x+\delta),y)].
\end{equation}
Formally expressed in Equation~\ref{eq:robust}, this objective aims to minimize the expected loss $L(\cdot)$ in the training data, subject to the maximization of the loss under perturbations bounded by a certain norm. Here, $L(\cdot)$ can be any suitable loss function, but typically the cross-entropy loss is used.
%and $L_{CE}$ is the cross entropy loss.
$\delta$ represents the adversarial perturbation applied to the input $x$ associated with the true label $y$. 
The term $||\cdot||_p$ denotes a norm $\ell_p$, commonly using $p=2$ for the Euclidean norm. 
The parameter $\epsilon$ ($epsilon$) restricts the magnitude of the perturbation within the specified norm. 
Inner maximization seeks the worst-case perturbation within the constraint set and is typically approximated using techniques such as projected gradient descent (PGD) \cite{madry2017towards}. 
PGD iteratively adjusts the perturbation in the direction of the loss gradient while ensuring it remains within the $epsilon$-bounded region. 
Adversarial training based on robust optimization aims to fortify models against adversarial perturbations, thereby improving their generalization and robustness in real-world scenarios. 
This approach has become pivotal in addressing security concerns in machine learning applications, especially in domains where model reliability is critical.

\subsection{Adversarial Loss (AdvL)}

In a bid to reconcile the robustness of models against adversarial attacks with their performance on clean data, the authors~\cite{kannan2018adversarial} followed the recommendations from~\cite{goodfellow2014explaining} and introduced a formulation that merges ERM with robust optimization that has a mixture of adversarial and clean data during training: 
\begin{equation} \label{eq:AdvL}
\min_{\theta} E_{(x,y)\sim D} [L(f_\theta(x),y) + \lambda L(f_\theta(x+\delta^*),y)].
\end{equation}
\begin{equation}\label{eq:delta}
      \text{where} \; \delta^* = \text{arg} \max_{||\delta||_p<\epsilon} L_{CE}(f_\theta(x+\delta),y).
\end{equation}
 AdvL strikes a balance between robustness and accuracy. The objective, as depicted in Equation~\ref{eq:AdvL}, minimizes the expected loss in the distribution training data $D$, incorporating both the standard cross-entropy loss and an augmented term. The augmented term, weighted by the hyperparameter $\lambda$, leverages robust optimization to encourage the model's resilience against adversarial perturbations. The adversarial perturbation $\delta^*$, calculated as the maximizer of the cross-entropy loss under a specified perturbation constraint ($||\delta||_p<\epsilon$) as presented in Equation \ref{eq:delta} (and typically solved for using PGD), guides the model to prioritize regions where its predictions remain stable under adversarial influence. By simultaneously optimizing accuracy and robustness, AdvL ensures that models maintain strong performance not only in clean settings, but also in the face of adversarial inputs.

\subsection{DataAug}

In \cite{Martínez-Martínez_Brown_Caceres_2024}, a modification is used from the AdvL formulation, but in this case instead of adversarial robustness, the target is distribution shift generalization:
%\begin{equation} \label{eq:DataAug}
%\begin{aligned}
%\min_{\theta} E_{(x,y)\sim D} \Big[ & L(f_\theta(x), y) \\
%& + \lambda L(f_\theta(g_{Aug=\text{ContrastShift}}(x)), y) \Big].
%\end{aligned}
%\end{equation}
\begin{equation} \label{eq:DataAug}
\begin{aligned}
\min_{\theta} E_{(x,y)\sim D} \Big[ & L(f_\theta(x), y) + \lambda L(f_\theta(g_{Aug=\text{ContrastShift}}(x)), y) \Big].
\end{aligned}
\end{equation}
The objective of Equation \ref{eq:DataAug} is to minimize the expected loss in the distribution training data $D$, incorporating both the standard cross-entropy loss and an augmented term. 
The augmented term, weighted by the hyperparameter $\lambda$, leverages generalization by including an additional sample $g_{Aug=ConstrastShift}(x)$ perturbed by a simple augmentation function. 
By simultaneously optimizing accuracy and generalization, DataAug encourages the model to maintain strong performance in clean data while being able to generalize to other distributions.

\subsection{AugMix}
In~\cite{hendrycks2019augmix}, a data enhancement strategy is used to improve the robustness and generalization of deep neural networks. The core idea revolves around the utilization of a chain of simple augmentation operations, which are applied stochastically and combined through a weighted sum to generate a diverse set of augmented images from a given input image. Using this augmentation strategy, the model is exposed to a broader spectrum of variations in the input data, thereby facilitating better learning of invariant features and improving its ability to generalize to unseen data. However, while augmentation can effectively increase the diversity of the training data, it also introduces challenges in maintaining consistency in the model's predictions across different augmented versions of the same input image. To address this issue, AugMix developers propose the use of Jensen-Shannon divergence (JSD) as a consistency loss~\cite{hendrycks2019augmix}. The AugMix training objective is as follows:
%\begin{align}
%\min_{\theta} E_{(x,y)\sim D} \big[ & L_{CE} (f_\theta(x), y) \notag \\
%& + \lambda \cdot L_{JSD} \big(f_\theta(x), f_\theta(g_{Aug}(x)), \notag \\
%& \hspace{20pt} f_\theta(g_{Aug}(x)) \big) \big],
%\end{align}
\begin{align}
\min_{\theta} E_{(x,y)\sim D} \Big[ & L_{CE} (f_\theta(x), y) + \lambda L_{JSD} \Big( f_\theta(x), f_\theta(g_{Aug}(x)), f_\theta(g_{Aug}(x)) \Big) \Big],
\end{align}
where, $L_{CE}$ is the cross-entropy loss, $L_{JSD}$ is the JSD loss, $g_{Aug}(x)$ is the stochastic function to apply a chain of augmentations and $\lambda$ is a scalar to help balance the contributions of the two loss terms. The JSD term is calculated by first computing the mean distribution $M$:
\begin{equation}\label{eq:M}
    M = \frac{{f_\theta(x) + f_\theta(g_{Aug}(x)) + f_\theta(g_{Aug}(x))}}{3},
\end{equation}
then, $M$ is used to calculate the JSD:
%\begin{align}\label{eq:jsKL}
%    \text{JSD} = \frac{1}{3} \Big( & \, \text{KL}[f_\theta(x) \| M] \notag \\
%    & + \text{KL}[f_\theta(g_{Aug}(x)) \| M] \notag \\
%    & + \text{KL}[f_\theta(g_{Aug}(x)) \| M] \Big),
%\end{align}
\begin{align}\label{eq:jsKL}
    \text{JSD} = \frac{1}{3} \Big( & \, \text{KL}[f_\theta(x) \| M] + \text{KL}[f_\theta(g_{Aug}(x)) \| M] + \text{KL}[f_\theta(g_{Aug}(x)) \| M] \Big),
\end{align}
where $KL$ is the Kullback-Leibler (KL) divergence between two distributions.

JSD loss serves as a regularization term in the training objective, in addition to the standard cross-entropy loss ($L_{CE}$), and encourages the model to produce consistent predictions for semantically equivalent images by minimizing the discrepancy between the probability distributions of the model output on the original image and its augmented versions.

\subsection{RobustAugMix}
On its own, AugMix is not robust to adversarial examples, and adversarial training alone will not guarantee robustness to non-adversarial perturbation types, such as natural variations. To combine the benefits of AugMix and adversarial training, and to develop models that are robust to both natural and adversarial data corruptions,~\cite{martinezrobustaugmix} propose optimizing the following objective:
%\begin{align}\label{eq:robustaugmix}
%\min_{\theta} E_{(x,y)\sim D} \Big[ 
%    L_{CE}(f_\theta(x), y) 
%    & + \lambda \cdot L_{JSD} \Big( f_\theta(x), \notag \\
%    & \hspace{10pt} f_\theta(g_{Aug}(x)), \notag \\
%    & \hspace{10pt} f_\theta(x + \delta^*) \Big) \Big],
%\end{align}
\begin{align}\label{eq:robustaugmix}
\min_{\theta} E_{(x,y)\sim D} \Big[ 
    L_{CE}(f_\theta(x), y) 
    + \lambda L_{JSD} \Big( f_\theta(x),
    f_\theta(g_{Aug}(x)),
    f_\theta(x + \delta^*) \Big) \Big],
\end{align}
where the third input to the JSD loss term from AugMix is replaced by an adversarial example, $x+\delta^*$ (see Equation \ref{eq:delta}, typically solved via PGD). This substitution serves to improve the robustness of the model against adversarial attacks while maintaining its robustness to common corruptions. In~\cite{martinez2023robust}, the RobustAugMix methodology was successfully applied in the domain of medical imaging, demonstrating its efficacy in enhancing robustness while simultaneously improving clean accuracy, a departure from the typical behavior observed with AT. This formulation maintained robustness to noise and adversarial attacks on X-ray images. However, this method has the trade-off that it sacrifices some adversarial accuracy to maintain the high clean accuracy.

\subsection{Robust Training with Data Augmentation (RTDA)}
 Since RobustAugMix sacrifices adversarial accuracy to maintain high clean accuracy, in this work, we propose a modification to the RobustAugMix formulation with the aim of maintaining good clean performance and generalization, without sacrificing adversarial robustness. We refer to this modified approach as Robust Training with Data Augmentation (RTDA):
%\begin{align}
%\min_{\theta} E_{(x,y)\sim D} \Big[ 
%    & L_{CE}(f_\theta(x + \delta^*), y) \notag \\
%    & + \lambda \cdot L_{JSD} \big(
%        f_\theta(x), 
%        f_\theta(g_{Aug}(x)), \notag \\
%    & \hspace{40pt} f_\theta(x + \delta^*)
%    \big) \Big]. \notag \\
%\end{align}
\begin{align}
\min_{\theta} E_{(x,y)\sim D} \Big[ 
    & L_{CE}(f_\theta(x + \delta^*), y)
    + \lambda L_{JSD} \Big(
    f_\theta(x), 
    f_\theta(g_{Aug}(x)),
    f_\theta(x + \delta^*)
    \Big) \Big]. \notag \\
\end{align}

The modified objective function replaces the cross-entropy term from RobustAugMix, applying it to adversarial samples, $x+\delta^*$ (Equation \ref{eq:delta}), instead of clean inputs~\cite{hendrycks2019augmix}. Optimizing cross-entropy loss using adversarial samples ensures that the model directly improves its adversarial robustness by learning to classify challenging, perturbed inputs correctly, strengthening decision boundaries against attacks.
In RTDA, the augmentation function, $g_{Aug}(x)$, is restricted to AugMix augmentations as RobustAugMix.
Similarly to AT, AdvL and RobustAugMix, the adversarial perturbation $\delta^*$, constrained by an $\ell_{p}$ norm with a magnitude limit $\epsilon$, is approximated by PGD to capture an approximation of worst-case scenarios. 
The JSD term aligns predictions across clean, augmented, and adversarial inputs in a manner similar to RobustAugMix that aligns these distributions. 
The modified objective strengthens model robustness to adversarial perturbations without inhering this RobustAugMix limitation and enhances generalization in real-world applications, making it particularly suitable for fields like medical imaging, where resilience to natural variations and consistency are essential.

\section{Experimental Set-Up}
\subsection{Datasets}
In this study, we used three different medical datasets with different classification tasks and image resolutions:

\begin{enumerate} 
\item X-ray dataset presented previously by~\cite{cohen2020covidProspective}: This data set contains 6432 samples where 20\% are samples for testing, with images resized to a resolution of 225×225. The classification task in this scenario is respiratory disease detection, with three classes: COVID-19, Pneumonia, and Normal.
\item Mammogram dataset used previously by~\cite{junbai2022}: Comprising 9173 samples, we split the data into 80\% for training and 20\% for testing. The images have a resolution of 1024×1024. This scenario is a binary classification task for breast cancer detection.

\item Point-of-care ultrasound (POCUS) dataset presented previously by~\cite{born2021accelerating}: This dataset consists of 3119 frames from 195 ultrasound videos, split similarly into 80\% for training and 20\% for testing. The images are resized to 255×255. Similarly to the X-ray dataset, the classification task is for the same three respiratory diseases mentioned.

\end{enumerate}
\begin{figure}
\centering
\includegraphics[width=0.58\textwidth]{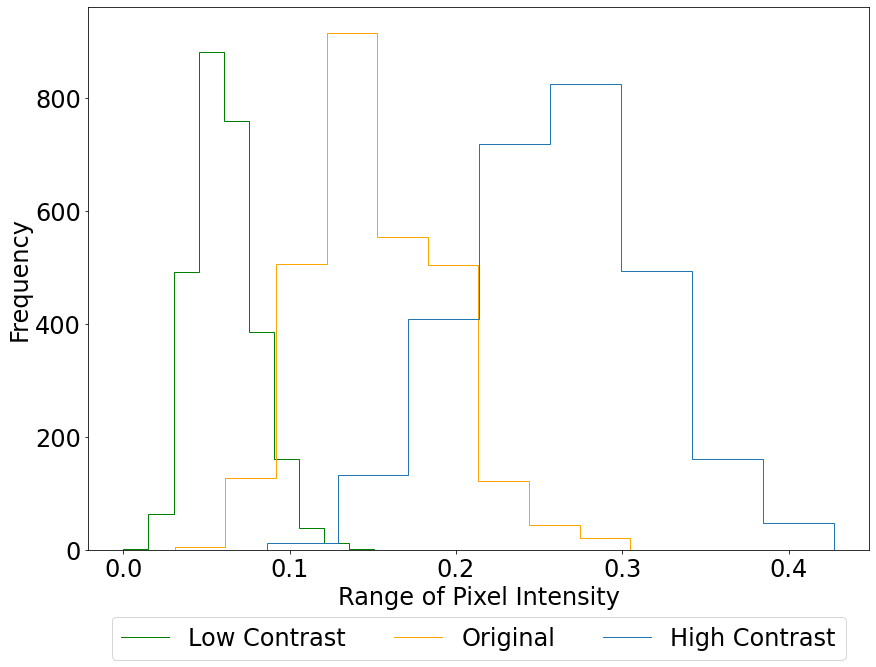}
\caption{Data distribution visualization before and after applying the natural variation perturbation on POCUS dataset.}
\label{fig:ultracontrast}
\end{figure}

Figure~\ref{fig:ultracontrast} presents the data distribution in the contrast domain before and after applying natural variations to the POCUS dataset. 
The high-contrast augmentation was applied to the X-ray and mammogram datasets as well during training and testing, but not the low-contrast since this is not a common scenario in these applications. 
Low contrast can occur with ultrasound probes such as the one presented by~\cite{Laura2023} if the practitioner overpressures the examination area.
During training, these augmentations are applied only for DataAug, AugMix, RobustAugMix and RTDA methods, but we will still evaluate all the methods against these distribution shifts. 
Furthermore, the adversarial samples are generated from the original distribution during training time only for AT, AdvL, RobustAugMix and RTDA methods, and for all the methods during testing.

\subsection{Models \& Parameters}

The models for the COVID-19 X-ray dataset were trained using ResNet18 architectures~\cite{deepResidual2016} for 40 epochs with a batch size of 32, using the SGD optimizer~\cite{ruder2017overviewgradientdescentoptimization} with a learning rate of 0.01. 
Similarly, models for the mammogram dataset used ResNet50 architectures trained for 50 epochs with a batch size of 5, using the same optimizer and learning rate. 
For the POCUS-COVID19 classification task, ResNet18 architectures were trained for 20 epochs with a batch size of 256, with an initial learning rate of 0.01 that decayed by a factor of 10 every 5 epochs. Models and hyperparameters are dataset dependent since they have different resolution, number of samples and data characteristics. The best hyperparameters were first found for the standard model, and then held fixed at those values for the robust methods, allowing us to focus our analysis on performance deviations from the standard model.

Across all datasets, five models per method were trained for cross-validation. 
The standard models were trained using the ERM formulation with cross-entropy loss.
The DataAug models were trained using cross-entropy loss and changes in contrast as the data augmentation.
For the methods that include adversarial examples within the training loop (AT, RobustAugMix, RTDA), PGD is used to solve for the adversarial perturbations using 7 gradient descent steps. PGD with 20 steps of gradient descent is used when testing all methods. 
Values for epsilon for PGD during training were dataset-specific: 2 for the COVID-19 X-ray dataset, 0.15 for the mammogram dataset, and 0.3 for the POCUS-COVID19 dataset. 
The epsilon value for the respiratory disease classification tasks was selected by evaluating the standard model and identifying where the adversarial accuracy starts to decrease.
In the case of the breast cancer classification task, we follow the study presented by \cite{joel2022using}, where the authors used a small epsilon value and achieved a more stable and robust model.
The Responsible AI Toolbox (RAI)~\cite{raitoolbox} was utilized to generate adversarial examples during training and testing.

\subsection{Metrics}

To ensure a rigorous evaluation of the performance of our method, we focused on two key metrics that provide complementary insights into its behavior in normal and challenging scenarios.
In this study, adversarial robustness was measured by calculating adversarial accuracy for different epsilon values.
This metric measures the proportion of correct predictions made by the model when confronted with adversarially perturbed samples, inputs specifically designed to exploit the model. 
Adversarial accuracy offers a stringent evaluation of the model ability to maintain performance under intentional perturbations. 

The adversarial accuracy is calculated as follows:
\begin{equation}
    \text{Adversarial Accuracy} = \frac{1}{N} \sum_{i=1}^N 1 \Big( f_\theta(x_i + {\delta_i}^*) = y_i \Big),
\end{equation}
where $N$ is the total number of samples in the dataset, ${\delta_i}^*$ is the adversarial perturbation of the sample, such as Equation \ref{eq:delta}, and $1(\cdot)$ is the indicator function, equal to 1 if the argument is true and 0 otherwise.

The Brier Score (BS)~\cite{brier1950verification} was used as a probabilistic metric to assess both the accuracy and the calibration of the model predictions. Specifically, we calculated the BS for both clean data and data subjected to distribution shifts. By evaluating the mean-square difference between predicted probabilities and true outcomes, this metric allowed us to quantify not only the correctness of the model's probabilistic outputs but also how well the predicted confidence levels align with the observed frequencies of outcomes. This dual evaluation on clean and shifted data provided critical insights into the model's ability to maintain reliable predictions when exposed to perturbations in the input distribution, reflecting its robustness in real-world deployment scenarios.

The BS is calculated as follow:

\begin{equation}
\text{BS} = \frac{1}{N} \sum_{i=1}^{N} (p_i - y_i)^2,
\end{equation}
where \(p_i\) is the predicted probability for the positive class, \(y_i \in \{0, 1\}\) is the true label, and \(N\) is the number of samples. A lower score indicates better calibration and reliability.

Together, these metrics provide a comprehensive assessment of the capabilities of the model. The BS offers a lens into its probabilistic soundness and calibration under both clean and shifted conditions, whereas adversarial accuracy examines its robustness to crafted perturbations. This combination ensures a nuanced understanding of the model's reliability and robustness across diverse operational scenarios, highlighting its readiness for deployment in both standard and adversarially challenging environments.

\section{Results}

Our results are presented using two complementary visualizations to effectively communicate adversarial robustness and distribution generalization. 
Adversarial robustness is presented using line graphs, where the x-axis represents the strength of the attack ($epsilon$), and the y-axis shows the mean adversarial accuracy of the model. Note that the left-most point on the x-axis ($epsilon=0$) represents the accuracy of the model on clean data.
For distribution generalization, we use grouped bar plots to display the BS results for each method on clean data (before the shift) and low/high contrast data (after the distribution shift), depending on the dataset.

\subsection{X-ray COVID-19}

\begin{figure*}[t]
\centering
\begin{minipage}{0.49\textwidth}
    \centering
    \includegraphics[width=\textwidth]{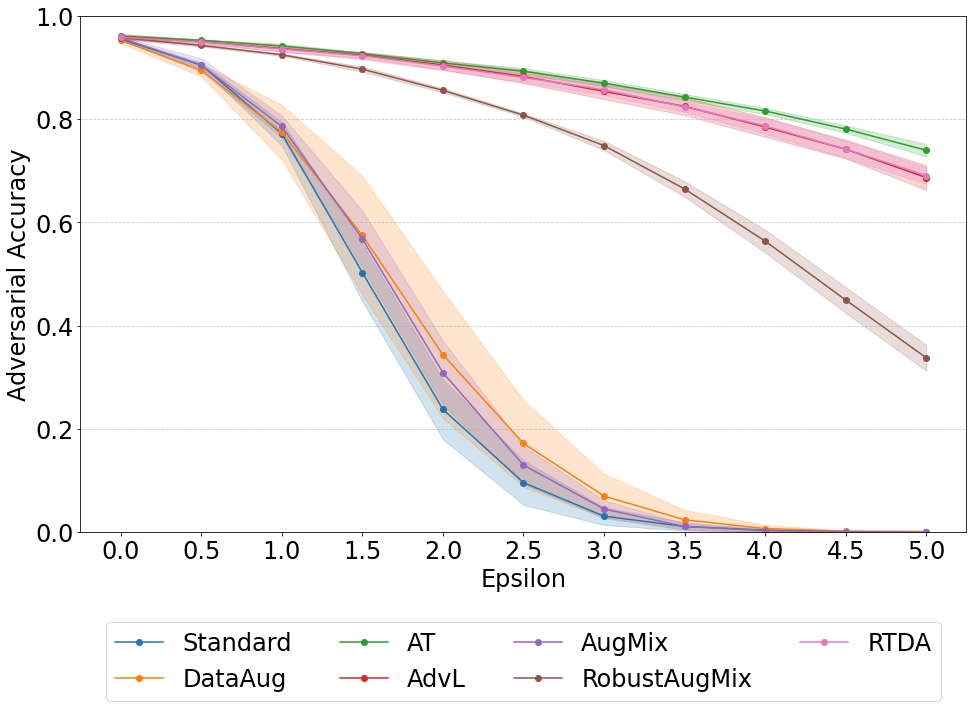}
    \subcaption{}
    \label{fig:xrayAdversarial}
\end{minipage}
\hfill
\begin{minipage}{0.49\textwidth}
    \centering
    \includegraphics[width=\textwidth]{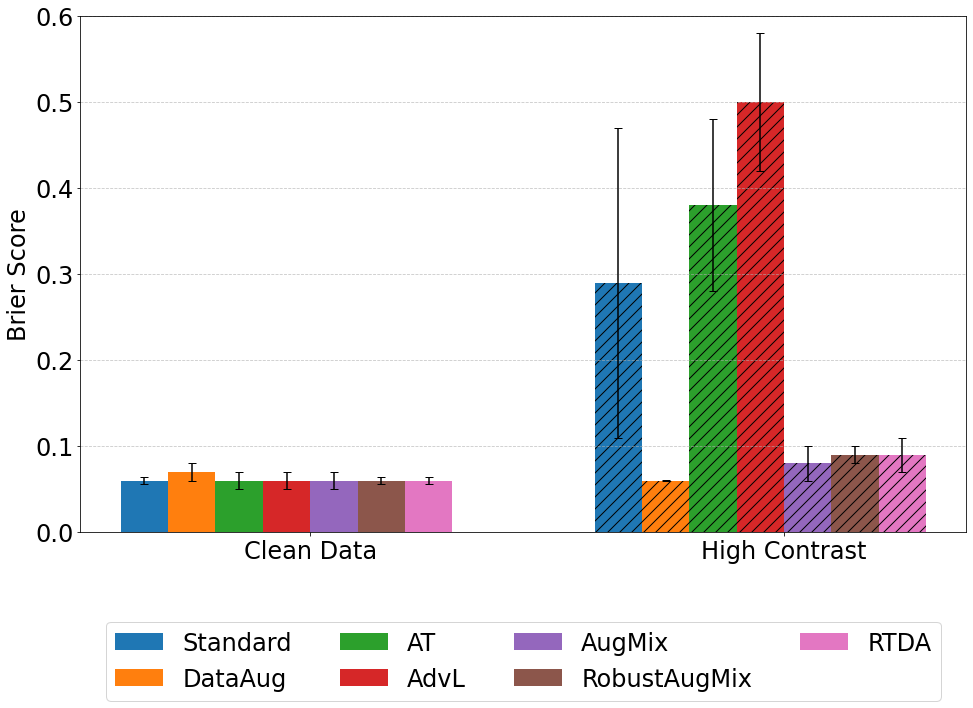}
    \subcaption{}
    \label{fig:xrayBrier}
\end{minipage}
\caption{Performance evaluation on the X-ray dataset: (a) Adversarial Accuracy (higher is better, shaded region around the lines denotes the standard deviation) and (b) Brier Score (lower is better, error bars denote the standard deviation).}
\label{fig:xrayPerformance}
\end{figure*}

Figure~\ref{fig:xrayAdversarial} highlights the performance of RTDA and the baseline methods with increasing adversarial perturbation strength to evaluate their adversarial robustness. 
The clean performance for these methods can be found in the figure where the value of $epsilon$ is 0.
RTDA demonstrates superior performance compared to other methods that can handle the distribution shift such as DataAug, AugMix and RobustAugMix, especially at higher epsilon values. 
The trend suggests that RTDA maintains a more gradual decline in accuracy as adversarial perturbations increase, indicating its robustness against stronger attacks. 
Among the methods evaluated, AT surpasses RTDA in terms of adversarial robustness at higher epsilon levels, but this margin of greater adversarial accuracy is limited to 3-5\%.
Interestingly, all methods that include adversarial examples in their training pipeline do not compromise their accuracy on clean data ($epsilon$ 0).

Figure~\ref{fig:xrayBrier} highlights the sensitivity of the X-ray classifiers to changes in distribution, especially when the model is not exposed to any data variations during training, such as the Standard, AT and AdvL methods. 
As expected, DataAug and AugMix were able to handle the distribution shift better than the rest of the methods.
It is expected since these two methods are trained explicitly to generalize to these other distributions, without any other additional requirements.
However, RTDA is the only method that maintains a low BS on both clean and shifted data while also achieving a high level of adversarial robustness. 

\subsection{Mammogram Breast Cancer}
\begin{figure*}[t]
\centering
\begin{minipage}{0.49\textwidth}
    \centering
    \includegraphics[width=\textwidth]{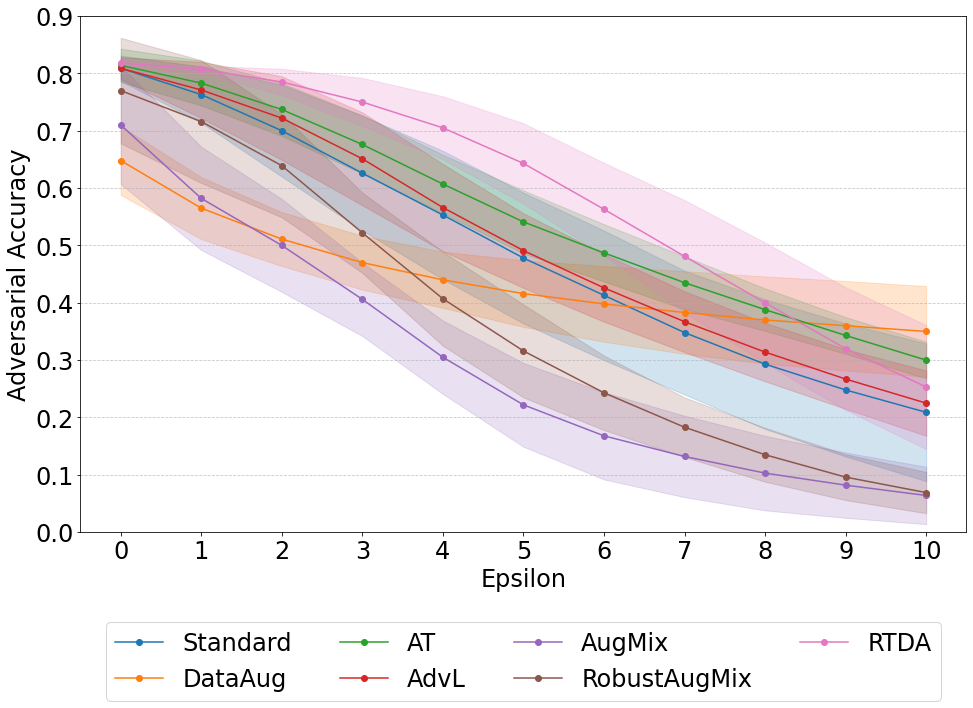}
    \subcaption{}
    \label{fig:mammAdversarial} % Subplot label for Adversarial Accuracy
\end{minipage}
\hfill
\begin{minipage}{0.49\textwidth}
    \centering
    \includegraphics[width=\textwidth]{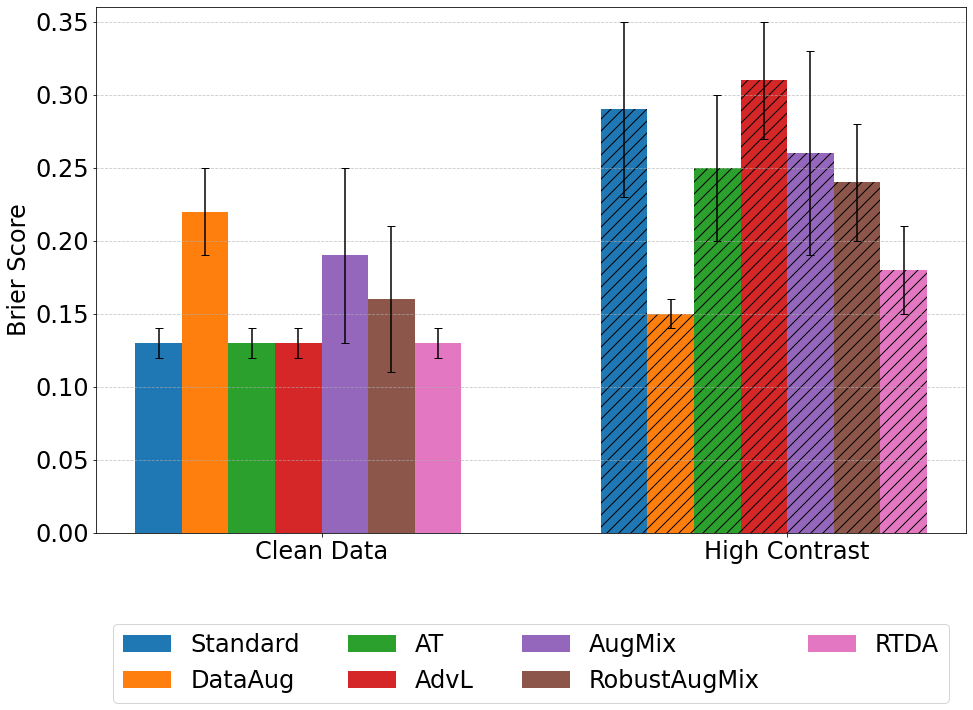}
    \subcaption{}
    \label{fig:mammBrier} % Subplot label for Brier Score
\end{minipage}
\caption{Performance evaluation on the mammogram dataset: (a) Adversarial Accuracy (higher is better, shaded region around the lines denotes the standard deviation) and (b) Brier Score (lower is better, error bars denote the standard deviation).}
\label{fig:mammPerformance} % Main figure label
\end{figure*}

Figure~\ref{fig:mammAdversarial} illustrates how the other methods besides of RTDA looking to generalize to other distributions are more vulnerable to adversarial attacks. 
RTDA consistently outperforms all other methods, including AT, at each $epsilon$ level up to the value of 8, after which it starts to drop and meets the other approaches with an accuracy below 40\%. 
Although the RobustAugMix loss function is conceptually similar, it fails to achieve the same level of adversarial robustness. 
This advantage of RTDA is due to its unique approach of solving for an adversarially perturbed image while simultaneously calibrating probabilities with clean and high-contrast samples. 
This strategy not only improves the resilience of the model against adversarial perturbations but also promotes the learning of essential features, leading to improved accuracy in clean samples ($epsilon$ 0). 
We note that surprisingly, using a small $epsilon$ value of 0.15 for this dataset during training achieved adversarial robustness to much larger values during testing. 
Also, even the standard model does not decrease performance below 50\% until $epsilon$ of 5, indicating it has some inherent robustness.
Another interesting point is that this is the only dataset in which AT is not the best performer on adversarial accuracy and is only more robust than the standard model by no more than 5\% in most of the $epsilon$ values.
Our hypothesis for these results is that since the mammogram images are high resolution (1024x1024), it is harder to craft adversarial attacks for models trained on this dataset.

Figure \ref{fig:mammBrier} shows how vulnerable this classification task is to a distribution change. 
DataAug demonstrates superior performance in the high-contrast domain by emphasizing high-contrast batch samples during training over cleaner ones.
The expectation was that AugMix would perform similarly to DataAug, since it actually sees two high-contrast samples during training in addition to the clean sample. 
RTDA achieved a lower BS on clean data than any other of the generalizable methods. 
Moreover, in the high-contrast distribution, only DataAug outperforms RTDA, indicating that RTDA is capable of generalizing effectively without compromising performance on clean data, unlike DataAug, AugMix, and RobustAugMix, which trade off clean performance for better generalization across distributions.

\subsection{POCUS COVID-19}
\begin{figure*}[t]
\centering
\begin{minipage}{0.49\textwidth}
    \centering
    \includegraphics[width=\textwidth]{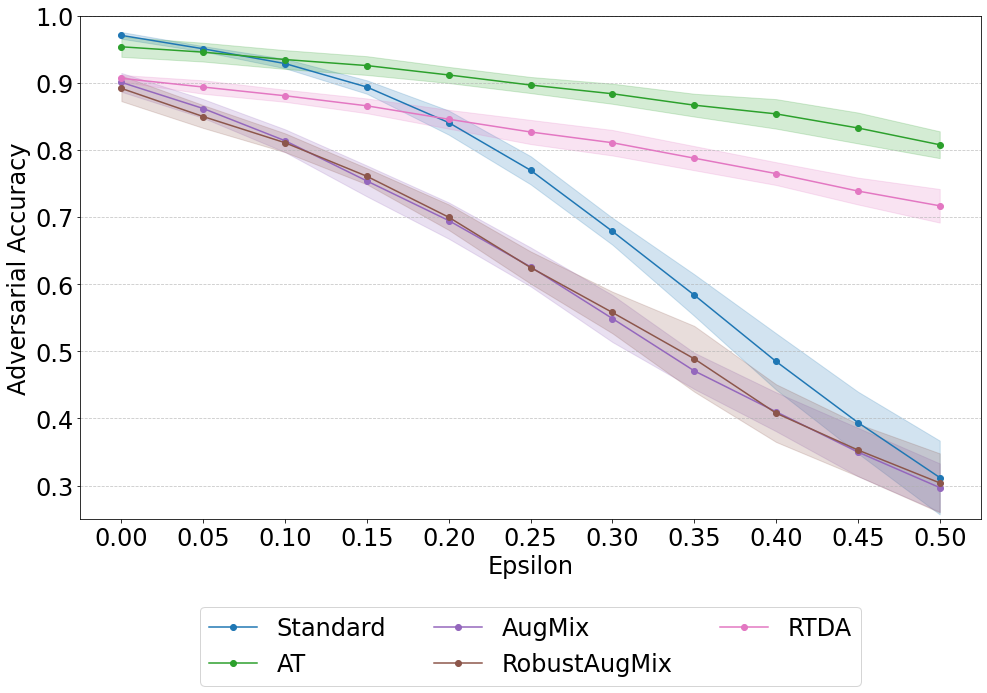}
    \subcaption{}
    \label{fig:ultraAdversarial}
\end{minipage}
\hfill
\begin{minipage}{0.49\textwidth}
    \centering
    \includegraphics[width=\textwidth]{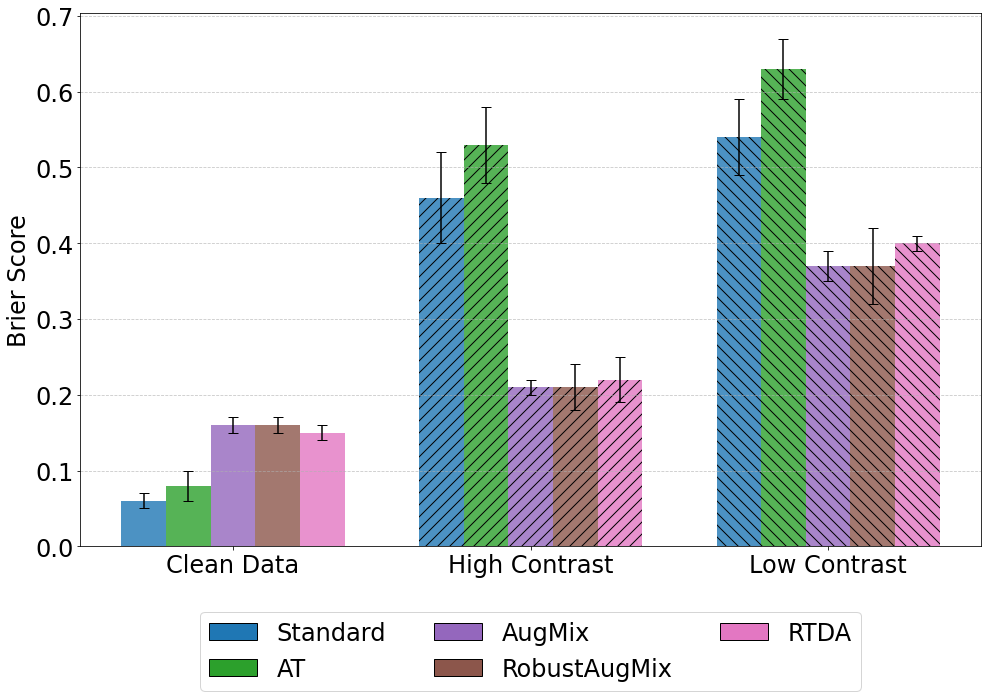}
    \subcaption{}
    \label{fig:ultraBrier}
\end{minipage}
\caption{Performance evaluation on the ultrasound dataset: (a) Adversarial Accuracy (higher is better, shaded region around the lines denotes the standard deviation) and (b) Brier Score (lower is better, error bars denote the standard deviation).}
\label{fig:ultraPerformance}
\end{figure*}

Figure~\ref{fig:ultraAdversarial} presents an interesting result, and it is that models that try to achieve adversarial robustness and generalize to the other two distributions suffer a decrease in clean and adversarial performance for this ultrasound dataset. 
Although RTDA is very stable up to the last $epsilon$ value tested, it never outperforms AT on adversarial robustness. 
However, even in the last $epsilon$ value, the RTDA accuracy is still above 70\%, which means that the model is much more robust compared to the other methods, such as RobustAugMix, which dropped to adversarial accuracy around 30\%. Furthermore, unlike the other two previous datasets, AT has compromised the clean performance of the POCUS COVID-19 classification pipeline.

Figure~\ref{fig:ultraBrier} presents how sensitive the standard and AT models can be to distribution shift. 
In this scenario, we are seeing an additional distribution, which is low-contrast. 
To generalize to this additional distribution we had to generate an additional augmented sample with low-contrast and add their probability distributions to the JSD loss.
RTDA is able to achieve lower BS than AugMix and RobustAugMix on clean data, however, those methods achieved a better performance under distribution shift. 
While RTDA has a higher BS on low-contrast, it has the smallest standard deviation of all the methods.
These comparisons demonstrate that RTDA is able to generalize to more than two distributions while maintaining adversarial robustness, but the trade-off is some sacrifice in clean accuracy ($epsilon$ 0).
DataAug and AdvL are not included in the POCUS classification task results since both methods suffered overfitting in this dataset.

\section{Conclusion}
In this study, we proposed RTDA, an approach to achieve robustness to adversarial attacks and natural variations that can cause distribution shift on medical imaging. 
The results demonstrate that RTDA achieves an impressive balance in the two evaluation metrics presented. 
Typically, one would expect standard training to excel in clean performance, DataAug or AugMix to be optimal for handling distribution shifts, and AT or AdvL to lead in adversarial robustness. 
However, RTDA emerges as the best or second-best performer in the evaluations under adversarial attacks and distribution shift, providing robustness without sacrificing clean performance in X-rays and mammograms, but sacrificing a small amount of clean performance in ultrasound. 
In particular, RTDA often achieved second-best performance under distribution shift, indicating strong resilience to natural variations while still effectively countering adversarial perturbations. 
RTDA has successfully balanced the need for high performance in standard settings with robustness to unexpected variations, making it a reliable model in diverse conditions. 
In future work, we plan to test RTDA against more types of adversarial attacks, train on other datasets with a larger number of samples, and incorporate real-world clinical validation to confirm the applicability of our method.

\section*{Acknowledgment}
The authors acknowledge partial support from the UConn Clinical Research and Innovation Seed Program (CRISP), led by Principal Investigator Sheida Nabavi, and the NASA Connecticut Space Grant Consortium. We also thank Bingjun Li and Derek Aguiar for their invaluable feedback, which significantly improved this work.

\end{document}